\documentclass{elsart}
\usepackage{graphicx}
\usepackage{bm} 
\usepackage{amssymb,amsmath,color,cite,multicol}
\usepackage[isolatin]{inputenc}
\usepackage[T1]{fontenc}
\usepackage{hyperref}

\journal{Physics Letters B}

\newcommand{\fsi}{FSI }

\begin{document}
\bibliographystyle{elsart-num}
\begin{frontmatter}
\title{Cross sections of the $pp\to K^+\Sigma^+n$ reaction close to threshold}
\author{The HIRES Collaboration:}
\author[c]{A.~Budzanowski},
\author[z]{A.~Chatterjee},
\author[o]{H.~Clement}
\author[o]{E.~Dorochkevitch},
\author[e]{P.~Hawranek},
\author[d] {F.~Hinterberger},
\author[d]{R.~Jahn},
\author[d]{R.~Joosten}
\author[f,w]{K.~Kilian},
\author[c]{S.~Kliczewski},
\author[f,w,m]{Da.~Kirillov},
\author[g]{Di.~Kirillov},
\author[h]{D.~Kolev},
\author[i]{M.~Kravcikova},
\author[e,f,w]{M.~Lesiak},
\author[f,w,m]{H.~Machner}\corauth[cor]{Corresponding author}
\ead{h.machner@fz-juelich.de},
\author[e]{A.~Magiera},
\author[l]{G.~Martinska},
\author[g]{N.~Piskunov},
\author[f,w]{D.~Proti\'c},
\author[f,w]{J.~Ritman},
\author[f,w]{P.~von Rossen},
\author[z]{B.~J.~Roy},
\author[w,d]{A.~Sibirtsev},
\author[g]{I.~Sitnik},
\author[c,d]{R.~Siudak},
\author[h]{R.~Tsenov},
\author[d]{K.~Ulbrich}
\author[l]{J.~Urban},
\author[o]{G. J.~Wagner}
\address[d]{Helmholtz-Institut f\"{u}r Strahlen- und Kernphysik der Universit\"{a}t Bonn, Bonn, Germany}
\address[g]{Laboratory for High Energies, JINR Dubna, Russia}
\address[m]{Fachbereich Physik, Universit\"{a}t Duisburg-Essen, Duisburg, Germany}
\address[f]{Institut f\"{u}r Kernphysik, Forschungszentrum J\"{u}lich, J\"{u}lich,
Germany}%
\address[w]{J\"{u}lich Centre for Hadron Physics, Forschungszentrum J\"{u}lich, J\"{u}lich, Germany}
\address[i]{Technical University Kosice, Kosice, Slovakia}
\address[l]{P.J.~Safarik University, Kosice, Slovakia}
\address[c]{Institute of Nuclear Physics, Polish Academy of Sciences, Krakow, Poland}
\address[e]{Institute of Physics, Jagellonian University, Krak\'{o}w, Poland}
\address[z]{Nuclear Physics Division, BARC, Mumbai, India}
\address[h]{Physics Faculty, University of Sofia, Sofia, Bulgaria}
\address[o]{Physikalisches Institut. Universit\"{a}t T\"{u}bingen, Germany}
%
%
\begin{abstract}
We have measured inclusive data on $K^+$-meson production in $pp$ collisions at COSY J\"{u}lich close to the hyperon production threshold and determined the hyperon-nucleon invariant mass spectra. The spectra were decomposed into three parts: $\Lambda p$, $\Sigma^0p$ and $\Sigma^+n$. The cross section for the $\Sigma^+n$ channel was found to be much smaller than a previous measurement in that excess energy region. The data together with previous results at higher energies are compatible with a phase space dependence.
\end{abstract}
\begin{keyword}
Meson production; hyperon-nucleon interaction
\PACS 13.75.-n, 13.75.Ev
\end{keyword}
\end{frontmatter}
%
%
\section{Introduction}\label{sec:Intro}

Hyperon--nucleon interaction has been of great interest from both the theoretical and experimental perspective. First, its knowledge is important for the understanding of hypernuclei. This requires the knowledge of the interactions in particular at small relative energy. Since hyperons are short lived this range is almost inaccessible by hyperon beams. An alternative is the study of final state interaction (FSI). In proton-proton interactions one has access to the reactions
\begin{eqnarray}
pp&\to &K^+\Lambda p\label{eqa:p-Lam}\\
pp &\to &K^+\Sigma^0p,\label{eqa:p-sig}\\
pp &\to &K^+\Sigma^+n \label{eqa:n-sig}.
\end{eqnarray}
For reactions (\ref{eqa:p-Lam}) and (\ref{eqa:p-sig}) there are several measurements of total cross sections in the threshold region \cite{Sewerin99, Kowina04, Abddesamad06, Balewski98, Bilger98, Valdau07}. Direct observation of $\Lambda p$ FSI comes from the energy dependence of the total cross section in reaction (\ref{eqa:p-Lam}). This cross section is enhanced close to threshold with respect to pure phase space behavior. In contrast, a similar enhancement was not observed in reaction (\ref{eqa:p-sig}), as is discussed in Ref. \cite{Sibirtsev06}. Much less is known for reaction (\ref{eqa:n-sig}). Two measurements exist (three points in total) from different groups at COSY which vastly disagree with each other. COSY-11 \cite{Rozek06} measured $K^+n$ coincidences while ANKE \cite{Valdau07} relied on  $K^+\pi^+$ coincidences with the pion coming from the decay $\Sigma^+\to\pi^+n$. The COSY-11 result indicates an enhancement of the cross section close to threshold, which might indicate a strong $\Sigma^+ n$ FSI. In this work we use a different method to determine the $pp \to K^+\Sigma^+n$ cross section in the threshold region.

The method chosen is based on the detection of the momentum spectrum of kaons. Then all three reactions contribute. The data can be converted into an invariant mass spectrum of the hyperon-nucleon system. The low mass part is due to the $K^+\Lambda p$ reaction. Above the $\Sigma N$ thresholds, 2128.9 MeV for the $\Sigma^+ n$ channel and 2130.9 MeV for the $\Sigma^0p$ channel, the yield is due to all three channels. We subtract the extrapolated $K^+\Lambda p$  cross section from this yield. The remaining double differential cross section represents the sum of the reactions $pp\to K^+ \Sigma^0 p$ and $pp\to K^+ \Sigma^+ n$. It can be used to deduce the sum of total cross sections for these two reactions.
Finally, subtracting the well known total cross section  for  $pp\to K^+ \Sigma^0 p$ yields the total cross section for $pp\to K^+ \Sigma^+ n$.

\section{Experiment and Results}

The experiment was performed at the COSY accelerator making use of the Big Karl magnetic spectrograph \cite{Drochner98, Bojowald02}. The tracks of charged kaons were measured under zero degree in the focal plane by stacks of multi-wire drift chambers. Particle identification was performed by time of flight measurement with scintillator walls following the drift chambers. Between these two detector arrangements we employed two threshold Cherenkov counters to veto events from pions \cite{Siudak08}. Kaon decay along the flight path was calculated with the CERN routine Turtle \cite{Turtle}. Finally, the count rates were converted into cross sections by making use of the measured incident beam intensity and target thickness \cite{Jaeckle94}. The experiments were performed at two beam momenta close to each other: 2.735 GeV/c and 2.870 GeV/c or kinetic energies of 1953.2 MeV and 2081.2 MeV, respectively. These correspond to excess energies of 133.2 MeV and 177.6 MeV for reaction (\ref{eqa:p-Lam}), 58.2 MeV and 102.6 MeV for reaction (\ref{eqa:n-sig}) and 56.3 MeV and 100.73 MeV for reaction (\ref{eqa:p-sig}). Further details can be found in Ref. \cite{Hires10}.

\begin{figure}[!h]
\begin{center}
\includegraphics[width=0.8\textwidth]{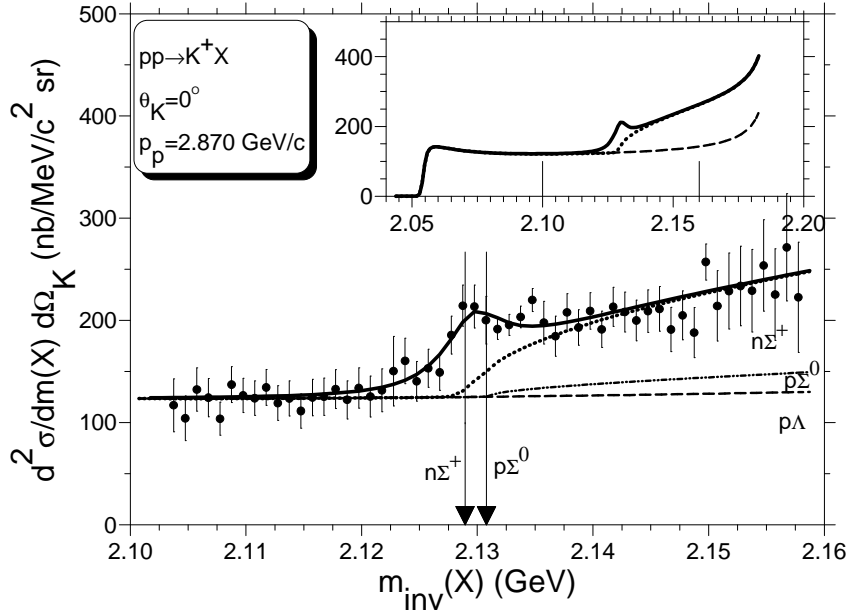}
\caption{Invariant mass spectrum of the hyperon-nucleon system for the incident beam momentum of 2870 MeV/c. The solid curve is the combined fit
for the three reactions $pp \to K^+\Lambda p$, $pp \to K^+ \Sigma^0 p$ and $pp\to K^+ \Sigma^+ n$.
The dashed curve is the extrapolation of the FSI enhancement for the $pp \to \Lambda p$ reaction assuming a  constant production matrix element.
The separate contributions to the three different reactions are also shown. For all calculations a Gaussian smoothing with $\sigma=$0.84 MeV was applied. In addition a Breit Wigner distribution is fitted to the cusp at the $\Sigma$ threshold. The insert shows the full inclusive kaon spectrum (solid curve) and the $pp \to K^+\Lambda p$ contribution (dashed curve).
The range of the present measurement is also indicated.  }
\label{Fig:Spectrum}
\end{center}
\end{figure}

Settings of the magnetic fields in the spectrograph were done in such a way that for the lower beam momentum only the region below the $\Sigma N$ threshold was measured while at the higher beam momentum the threshold region and above was measured. The data at the higher beam momentum are shown in Fig. \ref{Fig:Spectrum}.
Their origin will be be studied in more detail. We can express the cross section of the $\Lambda p$ interaction as:
\begin{equation}\label{equ:FSI_Bargmann}
\frac{d^2\sigma_\text{{lab}}}{d\Omega_{K}dm_{\Lambda{p}}}= \Phi_3 |\mathcal{A}|^2
\end{equation}
for a spin averaged final state. Here $\mathcal{A}$ is the amplitude which is given by \cite{Goldberger-Watson}
\begin{equation}\label{Eq:Amplitude}
|\mathcal{A}|^2=|{M}|^2 \, \frac{q^2+\beta^2}{q^2+\alpha^2}.
\end{equation}
$M$ is the production amplitude and the fraction is the final state interaction (\fsi)  enhancement, which is parameterized in terms of the Bargman potential parameters $\alpha$ and $\beta$, which can be transformed into scattering length and effective range.
With $q$ the internal momentum in the $\Lambda p$ system and
with $\Phi_3$ the three body phase space distribution normalized to the incident flux are denoted.
More details can be found in \cite{Hinterberger04}. The spectrum at the lower beam momentum was
then used to deduce
the effective range parameters of the $\Lambda p$ interaction \cite{Hires10}.
A three-parameter fit with spin-averaged effective range parameters
yielded a good description of the missing mass spectrum below the $\Sigma N$ thresholds. The deduced \fsi parameters were then applied also to the data taken at the higher beam momentum.
The resulting curve is shown together with the data in Fig. \ref{Fig:Spectrum}. It is extrapolated into the region above the $\Sigma^+ n$ threshold. A resonance like structure, which was also seen in exclusive $\Lambda p$ interaction following kaon absorption on the deuteron \cite{Tan69} was taken into account by replacing $|M|^2$ by $|M|^2+BW$ with $BW=A\Gamma^2/(Gamma^2+(m_{inv}-m_0)^2)$ a Breit Wigner distribution. The fit parameters are for the amplitude $A=57.5\pm7.0$ nb/MeV/c$^2$ sr, $m_0=2.1294\pm 0.0004$ GeV and $\Gamma=0.0030\pm 0.0005$ GeV for the case with smearing. When fitting without smearing the parameters change to $A=69.4\pm8.9$ nb/MeV/c$^2$ sr, $m_0=2.12934\pm 0.0003$ GeV and $\Gamma=0.0031\pm 0.0005$ GeV. However, a deeper discussion of this peak it is beyond the scope of the present letter, we plan to study this cusp in more detail in a separate paper.

In order to study whether this extrapolation is a valid procedure we performed several tests.   First we transform the double differential cross section  (\ref{equ:FSI_Bargmann}) from the laboratory system into the center of mass (c.m.) system and integrate over the full solid angle. This gives
\begin{equation}\label{Eq:CM}
\frac{d\sigma}{dm_{\Lambda{p}}} = \frac{1}{8(2\pi)^3}\frac{p^*_Kq}{p_p^*s}|\mathcal{A}|^2
\end{equation}
for the differential cross section.
Quantities with an asterisk are in the c.m. system and $s$ is the total energy squared. We then compare the prediction of Eq. (\ref{Eq:CM})
\begin{figure}[t]
\begin{center}
\includegraphics[width=0.5\textwidth]{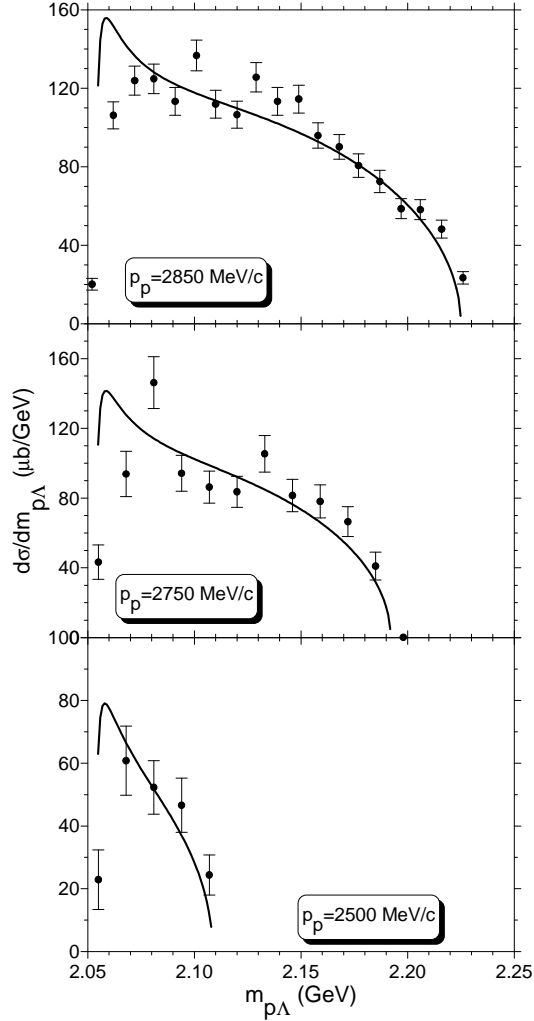}
\caption{Comparison of exclusive differential cross sections for $pp\to K^+\Lambda p$ production with calculations using Eq. (\ref{Eq:CM})  for beam momenta in the vicinity of the present ones. Data are from Refs. \cite{Bilger98, Metzger, Fritsch}.}
\label{Fig:Dalitz}
\end{center}
\end{figure}
with exclusive data (cf. Fig. \ref{Fig:Dalitz}). These were obtained by the TOF Collaboration \cite{Bilger98, Metzger, Fritsch} at beam momenta close to the present ones. The calculations contain only the\fsi parameters and the quoted total cross sections as input. There are no individually adjusted parameters. Apart from the \fsi enhancement near the $\Lambda p$ threshold there is good agreement between data and calculations.

We now study how valid is the procedure of integrating over $4\pi$. The validity requires isotropy of the kaon angular distribution. The partial wave analysis discussed further down indicates that the kaon is in a $s$ wave relative to the two baryon system. Moreover, Hogan et al. \cite{Hogan68} state, when inspecting their kaon angular distributions in the c.m. system, that "there does not seem to be any anisotropy". Their data span proton energies from 2.5 GeV to 3.0 GeV. More recent data from TOF \cite{Abdel-Bary10} taken at energies closer to energies employed in the present experiment (2.157 GeV to 2.396 GeV), show a similar behavior, and thus the total cross sections can be described as phase space times a \fsi.  Since our data are taken at smaller energies we add a 10$\%$ systematical uncertainty.

\begin{figure}[!h]
\begin{center}
\includegraphics[width=0.8\textwidth]{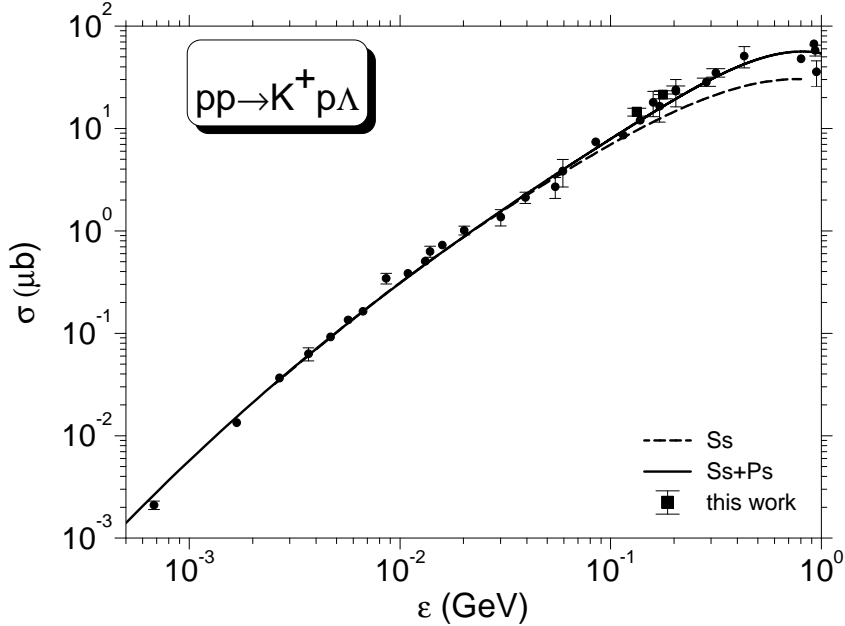}
\caption{Excitation function for the indicated reaction. The data shown by dots are direct measurements from Refs. \cite{Sewerin99, Kowina04, Abddesamad06, Balewski98, Bilger98, Valdau07, Flaminio84} and the very recent data from \cite{TOF2010}. The presently deduced total cross sections are shown by squares. The individual contributions of the partial waves $Ss$ (dashed curve) and $Ps$ as well as their sum (solid curve) are also shown.}
\label{Fig:Exfu_p_Lambda}
\end{center}
\end{figure}

Equation (\ref{Eq:CM}) can easily be integrated over the kinematical range yielding the total cross section. This implies of course that no orbital angular momentum is between the three particles, which is fulfilled close to threshold. We will discuss this assumption further down. The deduced values are shown in Fig.~\ref{Fig:Exfu_p_Lambda}.
The obtained total cross sections for the $\Lambda p$ channel are in good agreement with the world data \cite{Sewerin99, Kowina04, Abddesamad06, Balewski98, Bilger98, Valdau07, Flaminio84, TOF2010}.

Finally we give up the assumption of a constant average matrix element $M$. We apply the same procedure as in Ref. \cite{GEM02}. Again only three transitions were considered: ${^3P_0}\to{ ^1S_0s_0}$, ${^1S_0}\to{^3P_0s_0}$ and ${^3P_{0,1,2}}\to {^3P_{0,1,2}p_{0,1,2}}$. Here we have applied the common spectroscopic notation $^{2S+1}L_jl_J$ with $S$ and $j$ denoting the spin and
total angular momentum in the final $\Lambda p$ system,
respectively, and $J$ the total angular momentum. The capital symbol in the final state denotes the angular momentum between the two baryons and the small one the angular momentum of the kaon against this system. For $Ss$ again FSI was applied. In fitting the normalization constants it was found that two final states are sufficient to account for the data: $Ss$ and $Ps$. The resulting contributions for these two partial waves are shown in Fig. \ref{Fig:Exfu_p_Lambda}. As expected the $Ss$ contribution dominates at small energies. For the two beam momenta the fraction of $Ps$ is 17.2$\%$ and 21.3$\%$. In summary we find that the procedure applied is a sound method.

We proceed and subtract the extrapolated $pp \to K^+\Lambda p$ cross section from the data
in order to get the contribution from the reactions $pp\to K^+ \Sigma^+ n$ and $pp\to K^+ \Sigma^0 p$.
The cross section for the latter is obtained in the following way. We compile the world data for this reaction.
\begin{figure}[!h]
\begin{center}
\includegraphics[width=0.8\textwidth]{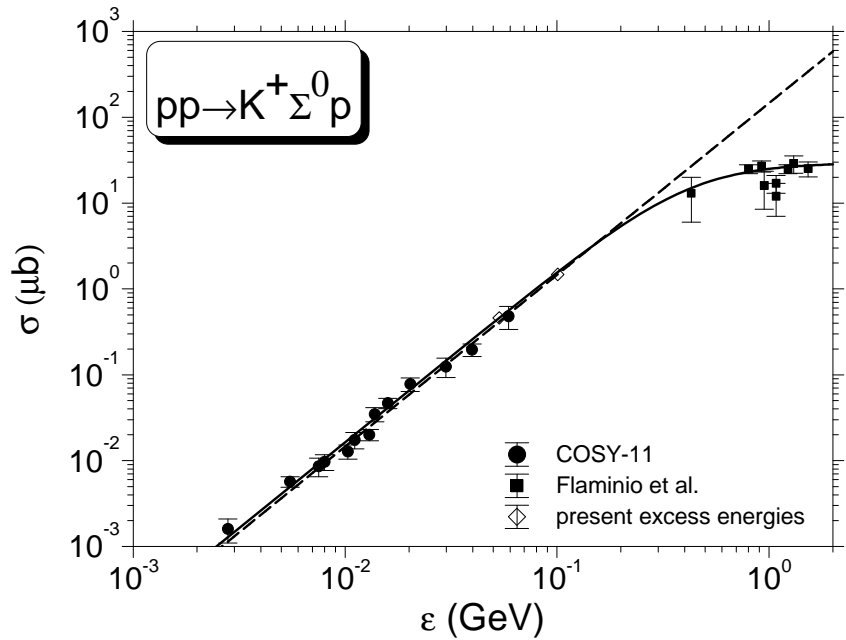}
\caption{Total cross sections for the indicated reaction as function of the excess energy $\epsilon$. The data shown as dots are from COSY-11 \cite{Sewerin99, Kowina04} and those shown as squares are older measurements compiled in \cite{Flaminio84}. The solid curve is a fit to the data, the dashed curve represents behavior of phase space. The deduced cross section for the two present excess energies are shown as open symbols.}
\label{Fig:Sigma_0}
\end{center}
\end{figure}
This is shown in Fig. \ref{Fig:Sigma_0}. We then fit a smooth curve to the data and compare this result to
the expectation for a uniform population of phase space. From this fit we extract the cross section for
the two beam energies. The two points obtained by this method are also shown in Fig. \ref{Fig:Sigma_0}. It is obvious that at the present excess energies one cannot distinguish between the pure phase space behavior and a smooth fit connecting the low energy data points with those at much higher excess energies. The double differential cross sections were then calculated from the total cross sections
by again making use of Eqs. (\ref{equ:FSI_Bargmann}) and (\ref{Eq:CM}) but without FSI  enhancement.
The obtained  $pp\to K^+ \Sigma^0 p$
cross section  is shown in Fig. \ref{Fig:Spectrum} on top of the
extrapolated $pp \to K^+\Lambda p$ cross section. This sum is then subtracted from the data yielding
the $pp \to K^+\Sigma^+ n$ cross section. The total sum of all three reaction cross sections is also shown in
Fig. \ref{Fig:Spectrum}. The same procedure as before is then applied to get the total cross section
for the reaction $pp\to K^+ \Sigma^+ n$.
The resulting total cross section is listed in Table \ref{Tab:1}
and shown in Fig. \ref{Fig:Sig_plus}.

The error is estimated using  the covariance matrix of the nonlinear least square fit. A possible systematic error of neglecting higher partial waves of $10\%$ is assumed.
\begin{figure}[!h]
\begin{center}
\includegraphics[width=0.8\textwidth]{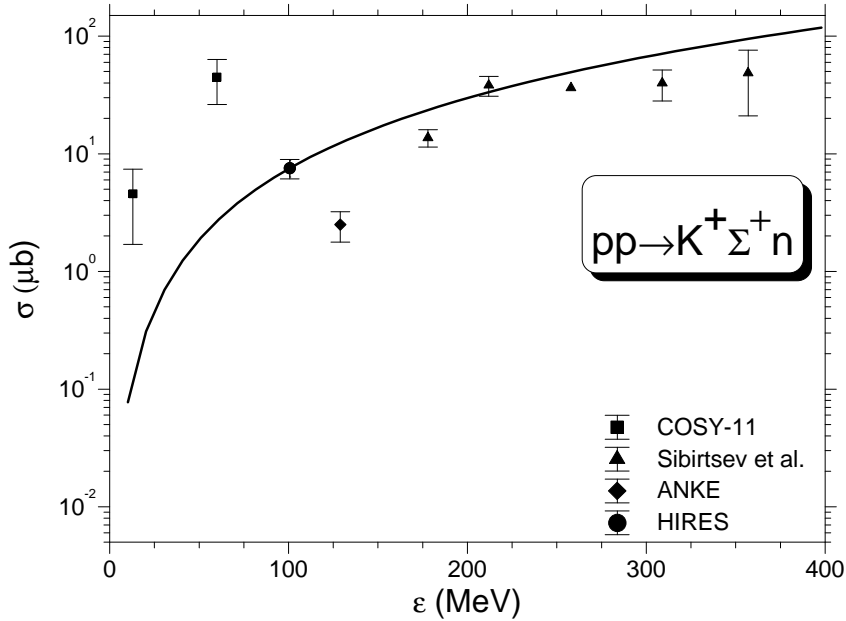}
\caption{The $pp\to K^+\Sigma^+n$ total cross section. The present result is shown as a full dot. The error bar contains
also the absolute normalization uncertainty added in quadrature.
The COSY-11 data \cite{Rozek06}
are shown as squares, the ANKE data point as a full diamond \cite{Valdau07}.
Cross section values reported by Sibirtsev et al. \cite{Sibirtsev07} are shown as triangles.
The solid curve is the phase space dependence normalized to the present data.}
\label{Fig:Sig_plus}
\end{center}
\end{figure}
\begin{table}
 \centering
\caption{ Total cross sections for the reactions $pp \to K^+\Lambda p$ and  $pp \to K^+\Sigma^+ n$
deduced from the missing mass spectrum shown in Fig. 1 via the method described in the text. The total cross section for the $K^+\Lambda p$ reaction at the lower beam momentum is deduced in a similar way from the data from Ref. \cite{Hires10}. The total cross sections for
the reaction $pp \to K^+\Sigma^0 p$ are deduced from the world data set shown in Fig. 3.
The indicated error of the  $pp \to K^+\Sigma^+ n$ cross section represents  the statistical plus systematical error.
In addition there is a 10$\%$ overall error due to the uncertainty in the luminosity calibration for the $pp \to K^+\Lambda p$ and  $pp \to K^+\Sigma^+ n$ reactions. The cross section for $pp \to K^+\Lambda p$ may have another systematical uncertainty of $10\%$ due to a possible anisotropy of the kaon angular distribution.} \label{Tab:1}
\begin{tabular}{|c|c|c|c|c|c|c|} \hline
 beam momentum  &\multicolumn{2} {c|} {$\Lambda p$} &\multicolumn{2} {|c|} { $\Sigma^+n$ }& \multicolumn{2} {|c|}{$\Sigma^0p$} \\ \hline
 (MeV/c)& {$\epsilon$ (MeV) }& {$\sigma$ ($\mu$b)} & { $\epsilon$ (MeV)} & {$\sigma$ ($\mu$b)}  &{$\epsilon$ (MeV)}&{$\sigma$ ($\mu$b)} \\ \hline
 2735 &133.2&14.5$\pm$1.3& 58.2&-  &56.3& 0.46$\pm$0.03 \\
 2870 &177.6&21.4$\pm$2.0& 102.6&7.5$\pm$1.2  &100.7 &1.48$\pm$0.09  \\ \hline
\end{tabular}
\end{table}

The present cross section is more than an order of magnitude smaller than the COSY-11 data and is slightly larger than the ANKE point. Sibirtsev et al. \cite{Sibirtsev07} performed
a similar analysis to extract the cross section from higher energy inclusive kaon spectra.
These results are also shown. We also indicate the energy dependence of the phase
space normalized to the present point. Obviously even the high energy data follow the phase space
dependence without needing to change the normalization constant.

\begin{figure}[h]
\begin{center}
\includegraphics[width=0.8\textwidth]{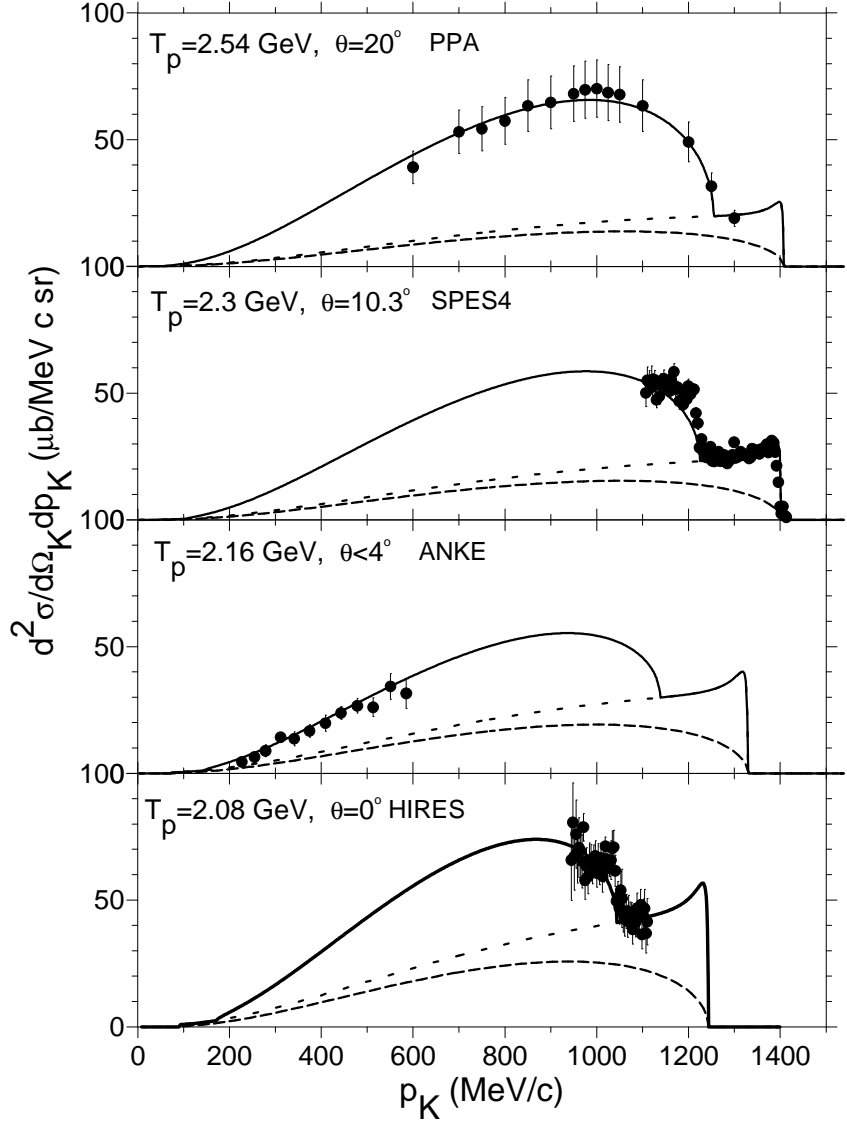}
\caption{Double differential kaon production cross sections vs. kaon lab momentum $p_K$ close to threshold. The data (dots) are the present results (HIRES), from  \cite{Valdau07} (ANKE),  \cite{Siebert94} (SPES4), and  \cite{Hogan68} (PPA). The adjusted cross sections for  $K^+\Lambda p$ with \fsi are shown as dotted curves, those ignoring \fsi as dashed curves. The solid curve is the sum of $K^+\Lambda p$ including \fsi and $K^+\Sigma N$.}
\label{Fig:Kaon_Spectra}
\end{center}
\end{figure}

Finally we compare the present data on inclusive kaon production in $pp$ interactions with
other data close to threshold \cite{Valdau07, Siebert94, Hogan68}. In Fig. \ref{Fig:Kaon_Spectra} we show the double differential cross sections as a
function of the laboratory momentum of the detected kaon.
The present data and those from \cite{Siebert94} cover a  limited range of  kaon momenta, but they are rather densely spaced.
The latter were transformed by us from invariant mass spectrum to kaon momentum spectrum. Those from \cite{Hogan68} cover a larger momentum range but are less densely spaced.
The data from \cite{Valdau07} are measured at rather low kaon momenta.  All data are compared with fits of $pp \to K^+\Lambda p$ cross sections with and without \fsi.
In all cases the \fsi parameters of \cite{Hires10} were applied. The matrix elements were adjusted to fit the region below the $\Sigma$ production thresholds.
For the data from \cite{Valdau07} we interpolate the production matrix element. Since for the smallest kaon momenta the difference between the calculations for $K^+\Lambda p$ with \fsi and without \fsi are minimal, \fsi parameters cannot be extracted from such a measurement. Furthermore very high resolution is required to see there the onset of $\Sigma$ production. For $\Sigma$ production we used for the present data the matrix elements from the present analysis, for those of Refs. \cite{Siebert94, Hogan68} we used the result from \cite{Sibirtsev07}. For the data of \cite{Valdau07}, the total cross section for the sum of the reactions $pp \to \Sigma^0 p$ and $pp \to \Sigma^+ n$ is about 9 $\mu$b.

\section{Summary}
We have measured inclusive kaon spectra at forward angles around $\theta=0^\circ$ with an missing mass resolution of $\sigma=0.84$ MeV for two beam momenta close to threshold. The data at the higher beam energy extend into the range of $\Sigma$ production. The quality of the data allowed us to subtract cross sections for reaction (\ref{eqa:p-Lam}) from the measured cross sections yielding the cross sections for the $\Sigma N$ channels.
The data close to threshold allow only for a very small \fsi for $\Sigma N$.
Total cross sections for these channels were obtained by integrating over the phase space volume.
Using a parametrization of the  $\Sigma^0 p$ channel cross sections the $\Sigma^+ n$ cross section is deduced. The phase space factor between these two reactions is five. A larger cross section for the charged $\Sigma$ is expected due to isospin considerations. The present result is slightly larger than the finding of ANKE but is lower by more than one order of magnitude than the COSY-11 data.

\bf{Note added in proof}

After finalizing the present study we became aware that the ANKE Collaboration has performed further measurements of the $K^+\Sigma^+n$ system closer to threshold
(Phys. Rev. C 81 (2010) 045208).

%
%
\section*{Acknowledgements}\label{sec:Acknowl}
We are grateful to the COSY crew for providing quality proton
beams. We appreciate
the support received from the European community research
infrastructure activity under the FP6 ``Structuring the European
Research Area'' programme, contract no.\ RII3-CT-2004-506078, from
the Indo-German bilateral agreement, from the Bundesministerium f\"{u}r Bildung und Forschung, BMBF (06BN108I), from the Research Centre
J\"{u}lich (FFE), and from GAS Slovakia (1/4010/07).
%



\end{document}